\documentclass[letterpaper,twocolumn,10pt]{article}
\usepackage{usenix-2020-09}

\usepackage{url}
\usepackage{subcaption}
\usepackage{enumitem}
\usepackage{graphicx}
\usepackage{tabularx}
\usepackage{caption}
\usepackage{titlesec}
\usepackage{multirow}
\usepackage{array}
\usepackage{booktabs}
\usepackage{pifont}
\usepackage{tikz}
\usepackage{xurl}
\usepackage{xspace}

\newcommand{\xmark}{\ding{55}}
\newcommand{\sysname}{StorageXTuner\xspace}

\usepackage[normalem]{ulem}

\newcommand{\myline}[1]{{\smallskip\noindent\textbf{#1.}}}
\newcommand{\mynewline}[1]{{\smallskip\noindent\textbf{#1}}}
\setlist{
  listparindent=\parindent,
  parsep=0pt,
}

\newenvironment{compactItemize}{\begin{list}{\scalebox{0.7}{$\bullet$}}{
    \setlength{\topsep}{0.8mm}
    \setlength{\itemsep}{0.25mm}
    \setlength{\parsep}{0.1mm}
    \setlength{\itemindent}{0mm}
    \setlength{\labelwidth}{2mm}
    \setlength{\labelsep}{2mm}
    \setlength{\leftmargin}{4mm}
    \setlength{\partopsep}{0mm}}
}{\end{list}}

\titlespacing*\section{0pt}{4pt plus 4pt minus 3pt}{4pt plus 2pt minus 0pt}
\titlespacing*\subsection{0pt}{3pt plus 3pt minus 3pt}{3pt plus 2pt minus 0pt}
\titlespacing*\subsubsection{0pt}{3pt plus 4pt minus 2pt}{1pt plus 2pt minus 1pt}
\begin{document}

\date{}

\title{\sysname: An LLM Agent-Driven Automatic Tuning Framework for Heterogeneous Storage Systems}

 \author{
 {Qi Lin$^1$, Zhenyu Zhang$^1$, Viraj Thakkar$^1$, Zhenjie Sun$^1$, Mai Zheng$^2$, Zhichao Cao$^1$}\\
 $^1$Arizona State University, $^2$Iowa State University}
\maketitle

\begin{abstract}
Automatically configuring storage systems is hard: parameter spaces are vast and conditions shift across workloads, deployments, and versions. Heuristic and ML tuners are usually tied to one system, rely on manual glue, and often lose effectiveness under changes. LLM-based proposals help, but most cast tuning as a single-shot, system-specific task, limiting cross-system reuse, constraining exploration, and offering weak validation. 

We present \textbf{\emph{\sysname}}, an LLM-agent-based auto-tuning framework for heterogeneous storage engines. \textbf{\emph{\sysname}} separates concerns across four agents—Executor (sandboxed benchmarking), Extractor (performance digest), Searcher (insight-guided configuration exploration), and Reflector (insight generation and management). The design couples an insight-driven tree search with a layered memory that promotes empirically validated insights and employs lightweight checkers to guard against unsafe actions. We implement a prototype and evaluate it on RocksDB, LevelDB, CacheLib, and MySQL InnoDB with YCSB, MixGraph, and TPC-H/C. Relative to out-of-the-box settings and to ELMo-Tune, \sysname reaches up to \textbf{575\%} and \textbf{111\%} higher throughput, reduces p99 latency by as much as \textbf{88\%} and \textbf{56\%}, and converges with fewer trials.

% Tuning storage systems is often challenging due to massive configuration spaces and workload/deployment variance. Existing heuristic- and machine learning-based solutions are typically system-specific, require human intervention, and fail on deployment changes. Recent Large Language Model (LLM) based approaches attack parts of this problem but usually treat tuning as a monolithic and system-specific task, leading to untapped cross-system generalizability, rigid iterative exploration, and a lack of robust validation mechanisms.

% This paper presents \textbf{\emph{\sysname}}, an LLM-agent-driven auto-tuning framework that operates across heterogeneous storage systems, workloads, deployments, and software versions. 
% To achieve this, \sysname introduces three key design innovations: 1) a modular agent architecture that decomposes tuning into benchmarking, analysis, configuration search, and insight management; 2) an insight-driven tree-based exploration strategy that prunes low-value trials while retaining cross-session knowledge; and 3) a layered memory design that leverages historically validated tuning insights. 
% Our prototype, evaluated on various storage systems, including RocksDB, LevelDB, CacheLib, and MySQL InnoDB under various workloads. \sysname improves throughput by up to \textbf{575\%} and \textbf{111\%} while reduces p99 latency by up to \textbf{88\%} and \textbf{56\%} over out-of-box and ELMo-Tune configurations, while converging faster with fewer trials. 

\end{abstract}

\section{Introduction}

Storage systems are widely used in modern IT ecosystems for data persistence and management. For instance, the social network infrastructures at Meta combine caching systems (e.g., TAO~\cite{tao}, Memcached~\cite{memcache}, CacheLib~\cite{berg2020cachelib}), graph stores (e.g., MyRocks~\cite{myrocks}, Neo4j \cite{neo4j_2025}, and Neptune\cite{neptune_2025}), key-value stores (ZippyDB~\cite{zippydb} and RocksDB \cite{rocksdb}), and object stores (e.g., S3\cite{amazon_s3_2025}, HDFS \cite{hdfs_site_2022, hdfs}) for various types of social data. Each of these storage systems provides numerous configuration parameters, and prior studies have demonstrated that tailored configuration tuning can substantially improve performance and resource utilization, often surpassing what default configurations can deliver~\cite{sehgal2010evaluating, carver-fast20, maricq2018taming}.

Tuning the storage system configuration typically follows a multi-stage pipeline. Initially, engineers need to understand targeted workload characterizations, upper-layer application behaviors, hardware specs, system performance metrics, as well as the software stack. Then, based on that information, engineers perform an iterative process of configuration exploration, evaluation, and re-adjustment \cite{10.14778/1687627.1687767}. Considering the hundreds of configuration parameters spanning diverse types with intricate dependencies~\cite{rocksdb, cachelibcode, innodb_mysql}, engineers usually use different algorithms like heuristic \cite{10.14778/1687627.1687767} or Bayesian \cite{powell_using_nodate} for tuning~\cite{kumar2011cost, tsialiamanis2012heuristics}. The new configurations will be evaluated via benchmarking, trace replay, or shadow testing for iterative adjustment until the expected tuning goal is satisfied \cite{10.14778/1920841.1920853}.

To achieve fast and efficient tuning of storage systems, different manual solutions are explored, including rule-based tuning \cite{cao2019practical} and search-based approaches like PAS~\cite{10.5555/1267359.1267363_provenance} and OASIS~\cite{xie2016oasis}. These methods typically rely on expert knowledge or guided exploration to navigate the configuration space. To automate the tuning process with less manual effort, Machine Learning (ML) is used in several auto-tuning studies, such as Endure \cite{Endure}, Dremel\cite{Dremel}, Ottertune~\cite{van_aken_inquiry_2021}, CDBTune~\cite{zhang_hbox_2021}, QTune~\cite{li_qtune_2019}. These methods build models that capture the relationship between configurations and performance metrics to automate the adjustments. More recently, several studies show that Large Language Modules (LLMs) hold strong promise for system tuning \cite{Thakkar2024can, thakkar2025elmo, giannankouris2024lambda, giannakouris2024demonstrating, lao_gptuner_2024}. ELMo-Tune~\cite{Thakkar2024can} applies prompt engineering to leverage LLM reasoning for key-value store tuning, overcoming the brittleness of expert-crafted heuristics. GPTuner~\cite{lao_gptuner_2024} and $\lambda$-Tune~\cite{giannankouris2024lambda, giannakouris2024demonstrating} use LLMs to guide database knob selection.

In general, LLM-based tuning solutions shift much of the tuning effort from handcrafted rules and system-specific models to automated reasoning~\cite{lao_gptuner_2024}, and are able to automate the whole tuning pipeline. Also, they show strong storage-system generality foundations with different system deployments ($\lambda$-Tune for Postgres and ELMo-Tune for RocksDB use the same underlying LLM model) and have the human-like reasoning ability to capture complex relationships across workloads, hardware, software, and configuration dependencies. 
Yet, current LLM-based tuning approaches remain limited. 
They are closely tied to specific system designs, which require substantial manual intervention and hard-coded engineering. 
Moreover, they typically collect all available information and feed it as a single LLM prompt, hoping that the LLM's reasoning capabilities can handle the tuning complexity and produce effective configuration recommendations. However, aggregating such heterogeneous information can weaken the LLM reasoning process~\cite{liu-etal-2024-lost}.
Further, these approaches inherit well-known drawbacks of LLMs: slow response, recurring API costs, and susceptibility to errors~\cite{chen2023frugalgptuselargelanguage}.

To overcome these limitations for achieving an efficient LLM-based storage system tuning solution, several key challenges must be addressed:
\uline{1)} Existing LLM-based tuning frameworks are tightly coupled to a specific storage system design or an implementation version. A key challenge is how to decouple system-specific semantics (e.g., system APIs, version-specific behaviors, and implementation-specific details) and reduce manual effort.
\uline{2)} LLMs prefer to work with clear, task-specific, and structured contexts\cite{liu2021pretrainpromptpredictsystematic, sun2025tablethoughtexploringstructured}. It is critical to divide the whole tuning pipeline into subtasks with clear scope and context, so that LLMs can focus on accurate reasoning and decision-making.
\uline{3)} Guiding LLMs to explore large configuration spaces with high efficiency is challenging. It is essential to explore search algorithms to avoid costly exhaustive searches and minimize the overhead of trial-and-error tuning.
And \uline{4)} given LLMs’ tendency to hallucinate and overfit to prompt phrasing \cite{Ji_2023,sclar2024quantifyinglanguagemodelssensitivity}, a key challenge is how to ensure LLM tuning suggestions are accurate and reliable.

To address the aforementioned challenges, we propose \sysname, an end-to-end, fully automated, and general storage system tuning framework that integrates LLM reasoning while ensuring efficiency and robustness. Inspired by the recent success of LLM agents, which extend LLMs with planning, tool use, and collaboration abilities \cite{yao2023reactsynergizingreasoningacting,wu2023autogenenablingnextgenllm}, \sysname introduces three key innovations:

\begin{compactItemize}
    \item \textbf{\textit{Collaborative Multi-Agent Tuning Framework}}:
    \sysname decomposes and abstracts the tuning process into four stages with dedicated LLM agents that operate in an iterative pipeline: 1) Executor is responsible for storage system deployment, benchmarking, and monitoring; 2) Extractor analyzes deployment, workload, benchmarking statistics, and log data collected by Executor and summarize them into structured tuning input; 3) Searcher explores the configuration space, decide the next round of configurations based on the input from Extractor and Reflector, and summarize the tuning experience; And 4) Reflector collect, analyze, and manage the tuning insights from the past tuning experiences from Searcher. 
    Moreover,  we design a set of agent-specific checkers to validate LLM outputs before execution to further enhance robustness.
    
    \item \textbf{\textit{Insight-Driven Exploration of Configuration Space}}: \sysname combines LLM reasoning with tree-based exploration to navigate complex configuration spaces. It extracts high-level tuning insights from past trials and injects them into the Searcher’s context, guiding the generation of candidate configurations as new tree branches. Searcher selects the most promising candidates from multiple branches for further expansion based on the analysis results from Extractor, efficiently focusing on high-potential configurations while avoiding redundant or suboptimal paths.
    \item \textbf{\textit{Memory-Efficient Management of Tuning Insights}}:
    \sysname generates tuning insights and manages them in a layered memory by Reflector: Short-Term Memory (STM) for tentative insights and Long-Term Memory (LTM) for validated insights. A confidence score is associated with each insight to reflect the reliability and is dynamically adjusted. Tuning insights are retrieved based on contextual similarity and confidence scores during the configuration searching process, enhancing LLM reasoning efficiency and adaptability for different workloads and environments.
\end{compactItemize}

We implement a prototype of \sysname with code released on GitHub\footnote{\url{https://github.com/gdaythcli/StorageXTuner} (Anonimized)}. We evaluate \sysname across various single-node storage systems (e.g., key-value stores and caches), including RocksDB \cite{rocksdb}, LevelDB \cite{leveldb}, CacheLib \cite{berg2020cachelib}, and MySQL InnoDB \cite{innodb_mysql}. We do not include distributed storage systems, as they introduce additional challenges such as network variability, coordination overhead, and fault tolerance, which are beyond the scope of this paper. 

\sysname delivers consistent performance improvements across multiple baselines, including default configurations, LLM-Default, and state-of-the-art tuners (ELMO-Tune, ADOC, Endure for RocksDB; SMAC, DDPG, and $\lambda$-Tune for SQL InnoDB). Real-world workloads such as YCSB \cite{ycsb}, MixGraph \cite{cao2020characterizing}, and TPC-H/C \cite{tpc-h} reveal throughput gains up to 575\% and p99 latency reductions up to 88\%. For CacheLib, \sysname improves hit ratios by up to 3.1 percentage points and achieves up to 22\% higher throughput compared to other baselines across write-intensive, read-intensive, and mixed workloads. In MySQL InnoDB, \sysname yields up to 709\% improvement in transaction throughput on TPC-C and 71\% improvement in query performance on TPC-H. We further evaluate \sysname’s flexibility across system versions (four RocksDB and two LevelDB releases), tree-search branching factors (child nodes), insight counts, and different LLMs (GPT-3.5, GPT-4o, GPT-o3-mini, and LLaMA variants), demonstrating robust performance under diverse conditions. 

In addition, we present our learnings from \sysname, presenting how the configuration search process majorly revolves around a few key parameters, why LLMs perform better when asked to edit instead of create configurations, and having a closed-loop LLM-reasoning architecture significantly reduces configuration convergence iterations.

\section{Background \& Motivations}

\subsection{Configuring and Tuning Storage Systems}

    Storage systems like caching systems (e.g., CacheLib \cite{cachelibcode}, Memcached \cite{anthonymemcached}, and Redis \cite{redis}), persistent key-value storage engines (e.g., RocksDB \cite{rocksdb}, LevelDB \cite{leveldb}, and MySQL InnoDB \cite{innodb_mysql}), and file systems (e.g., Ext4 \cite{ext4_man_2025}, XFS \cite{xfs_man_2025}, Tectonic \cite{tectonic}, and HDFS \cite{hdfs}) are important for various applications to store and manage a large volume of data. Out-of-the-box configurations usually fall short in meeting application requirements of performance metrics like throughput and latency in different scenarios \cite{sehgal2010evaluating}. Typically, each storage system exposes a number of tunable configurations (e.g., more than 170 in RocksDB, and 140 in MySQL InnoDB), which control diverse aspects such as I/O options, buffer sizes, threads, and other critical storage system behaviors. The large configurational space, along with the targeted workloads \cite{ycsb, cao_characterizing_2020, 2012workload, 2000workload} and hardware deployments \cite{zhang2021towards, dong2023disaggregating} in which these systems are used, makes tuning configurations a challenging task.

    Domain experts or engineers typically tune configurations \cite{wang2023characterizing} with a multi-step process. They profile targeted workloads into representative benchmarks (e.g., social graph OLTP workload from Meta as the MixGraph benchmark \cite{cao2020characterizing}) to enable iterative offline testing. 
    Then, the storage system configurations are iteratively tuned and evaluated via benchmarking to observe the impact on key performance metrics. During the tuning process, a number of factors, including storage system design, interplay of the workload profile (e.g., Zifian access pattern, read-heavy workloads), deployment setup (e.g., storage devices, CPU, and Memory), hardware and software characteristics, and resource constraints, are also considered. Once the experts achieve satisfactory performance as required by the application, the configurations are pushed for production deployment. 

    To automate the iterative tuning process, Machine Learning (ML) is used in several auto-tuning studies, such as Endure \cite{Endure}, Dremel\cite{Dremel}, Ottertune~\cite{van_aken_inquiry_2021}, CDBTune~\cite{zhang_hbox_2021}, QTune~\cite{li_qtune_2019}, CAPES \cite{li_capes_2017}, CAMAL \cite{yu2024camal}, ADSTS \cite{lu_adsts_2022}, and H5Tuner \cite{behzad_taming_2013_h5tuner}. These solutions usually focus on exploring parameters affecting performance and building an automated tuning framework that leverages learned patterns to tune the configurations. More recently, interest has grown in tuning solutions that are cohesive with Large Language Models (LLMs). 
For example, ELMo-Tune~\cite{Thakkar2024can} introduces an LLM-based framework for tuning RocksDB, a key-value store, across different workloads and deployment environments, $\lambda$-Tune~\cite{giannankouris2024lambda} demonstrates that LLMs can generate effective database configuration suggestions under diverse workloads, and GPTuner~\cite{lao_gptuner_2024} integrates LLMs with Bayesian optimization to improve configuration exploration efficiency. Those studies highlight new opportunities for addressing the complex task of storage system tuning with LLMs.

\subsection{LLMs and LLM Agents}
\myline{Large Language Models (LLMs)} 
Built upon transformer architectures, LLMs are advanced AI systems that recognize complex patterns in language and context, powering a wide range of NLP tasks such as translation, question answering, summarization, and content creation \cite{GPT4, openai_gpt-4_2024, gpt_4o_welcome_page}. Prominent examples include GPT \cite{openai_gpt-4_2024}, Gemini \cite{gemini_google_nodate}, Claude \cite{noauthor_claude_nodate}, and Copilot \cite{noauthor_copilot_nodate}. LLMs are pretrained on massive corpora, giving them a broad knowledge base and strong generalization capabilities across different domains \cite{gpt_4o_welcome_page, openai_gpt-4_2024, brown2020languagemodelsfewshotlearners}. The human-like reasoning helps to infer the purpose of configurations and relationships between configuration changes and performance \cite{chen2021evaluatinglargelanguagemodels}.

\myline{LLM Agents} 
Unlike standalone LLMs, LLM agents can integrate additional actions, access plugins, call APIs, and orchestrate complex workflows, effectively acting as intelligent assistants capable of more than natural language responses. Techniques such as Chain-of-Thought prompting~\cite{wei2023chainofthoughtpromptingelicitsreasoning} encourage the model to generate intermediate reasoning steps explicitly, enabling more accurate and explainable outputs. Tree-of-Thoughts~\cite{yao2023treethoughtsdeliberateproblem} allows the model to explore multiple branching reasoning paths in parallel, evaluating alternatives before committing to a solution. Self-refinement~\cite{madaan2023selfrefineiterativerefinementselffeedback} lets the model iteratively critique its own outputs and make improvements, improving reliability and reducing errors. These innovations have demonstrated success across domains such as software engineering, automation, optimization, question-answering, and information retrieval\cite{wang2025mllm, dong2025mmtok, wang2025llm, wang2025efficient, gong2025unsupervised, gong2025agentic, gong2025evolutionary}.

\mynewline{LLM: The New Silver Bullet?}
LLMs show strong potential for storage system tuning, offering generalization across heterogeneous workloads and deployments with minimal human intervention \cite{Thakkar2024can, thakkar2025elmo}. Pretrained on massive and diverse corpora~\cite{openai_gpt-4_2024, huggingface_llm_dataset}, LLMs already have rich knowledge of most of the open-source storage systems and can reason across multiple dimensions of tuning complexity, such as workload diversity, hardware constraints, software behaviors, and configuration dependencies~\cite{chen2021evaluatinglargelanguagemodels}. LLMs can map high-level insights, like “increasing cache size improves read performance,” to system-specific actions such as adjusting \textit{block\_cache\_size} in RocksDB or \textit{innodb\_buffer\_pool\_size} in MySQL.

\subsection{Limitations of State-of-The-Art}
Despite their potential, current LLM-based solutions for storage system tuning remain limited. First, existing LLM-based solutions typically rely on a single LLM inference cycle for narrow subtasks, such as directly generating configuration suggestions from workload statistics~\cite{Thakkar2024can}. However, a single LLM struggles with the multi-stage nature of storage system tuning, which involves workload characterization, performance diagnosis, configuration search, and validation. Feeding heterogeneous information all at once often overwhelms the model and degrades its reasoning~\cite{liu-etal-2024-lost}. To illustrate this, we compare two tuning approaches for RocksDB using the Elmo-Tune framework:
\begin{compactItemize}
\item \textbf{Single-shot input}: feeding all relevant information (e.g., hardware, software, configuration parameters, and workload characteristics) into a single LLM at one prompt.
\item  \textbf{Multi-shot input}: dividing the same information into semantically coherent batches and sending each batch to separate LLM agents. Each agent processes its context independently, and their outputs are then integrated to produce final tuning recommendations.
\end{compactItemize} 
To ensure a fair comparison, both approaches use the same input content and information. Under the MixGraph workload\cite{cao_characterizing_2020} in RocksDB, the multi-shot approach significantly outperforms the single-shot method (as shown in Table~\ref{tab:combined-comparison}), highlighting the advantage of structured, focused reasoning for multi-stage storage system tuning.

Second, existing LLM-based approaches lack mechanisms to efficiently explore large and complex configuration spaces, often resulting in costly trial-and-error searches. For instance, Table~\ref{tab:combined-comparison} shows that under the MixGraph workload in RocksDB using the Elmo-Tune framework, random search (the default in ELMo-Tune) incurs substantially higher exploration overhead than using tree search in Elmo-Tune when achieving the same performance level. This occurs because ELMo-Tune relies on a single LLM inference cycle per suggestion, which causes it to repeatedly evaluate the same or similar settings before converging on effective configurations.

\begin{table}[]
    \footnotesize
    \centering
    \setlength{\tabcolsep}{0.475em}
    \caption{Comparison of Input Types and Search Methods.}
    \label{tab:combined-comparison}
    
    \begin{tabular}{clcc}
        \toprule
        \multirow{3}{*}{\textbf{\begin{tabular}[c]{@{}c@{}}Input \\ Type\end{tabular}}} & \textbf{Strategy} & \textbf{Throughput (kops/s)} & \textbf{Latency ($\mu$s)} \\
        \cmidrule(lr){2-4}
                     & Single-shot input & 104 & 78 \\
                    & Multi-shot input  & 231 & 31 \\
        \midrule
        \multirow{3}{*}{\textbf{\begin{tabular}[c]{@{}c@{}}Search \\ Methods\end{tabular}}}            & \textbf{Strategy} & \textbf{Total Token (K)} & \textbf{Per-Round (K)} \\
        \cmidrule(lr){2-4}
                     & Random search & 142  & 6.76 \\
                    & Tree search   & 63   & 12.6 \\
        \bottomrule
    \end{tabular}
\end{table}

Furthermore, existing LLM-based approaches \cite{Thakkar2024can, giannankouris2024lambda, lao_gptuner_2024} tightly couple system-specific semantics with the tuning framework, leading to low generality and compatibility. For example, Elmo-Tune~\cite{Thakkar2024can} requires developers to manually hard-code RocksDB interfaces and modify the tuning framework source code to apply LLM-generated suggestions. Moreover, Elmo-Tune cannot successfully tune different RocksDB versions directly. Through careful testing across different RocksDB versions, we found that ELMo-Tune works correctly on version 8.8.1, but fails on version 5.7.1 due to version compatibility issues. Therefore, existing LLM-based approaches are neither fully automated nor general, and they still demand substantial human effort to ensure compatibility with particular implementations or release versions.

\begin{figure*}
    \centering
    \includegraphics[width=0.9\textwidth]{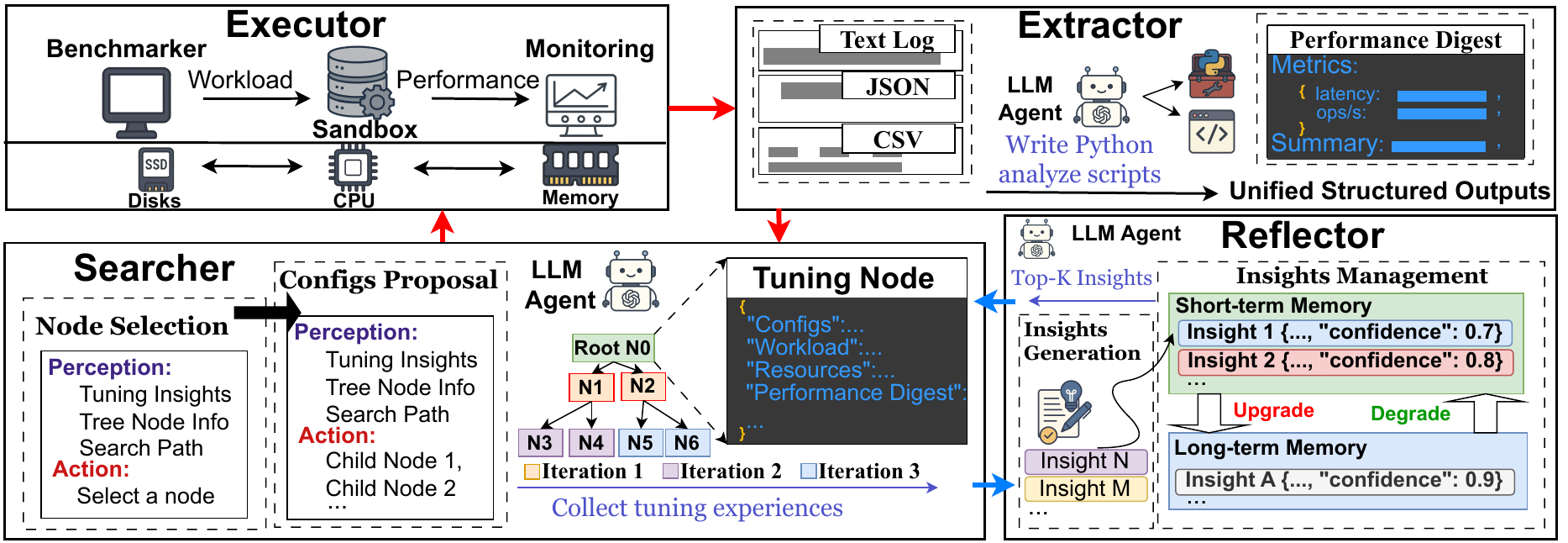}
    \caption{\sysname framework 
    } 
    \label{fig:arc}
\end{figure*}

Additionally, current approaches cannot retain or reuse the tuning knowledge from previous tuning cycles or even cross different tuning sessions (but engineers will learn from those past trials). In other words, even if a workload has been tuned before, ELMo-Tune treats each new tuning session fully independently, requiring repeated exploration of the configuration space from scratch.
Table~\ref{tab:workload_performance} shows that ELMo-Tune with prior tuning knowledge (e.g., historical configurations and their observed performance) effectively reduces the number of exploration iterations and achieves better RocksDB performance than the default ELMo-Tune across different workloads.

Together, these limitations motivate the need for \emph{a holistic, automatic storage system tuning framework that integrates LLM reasoning across all stages of the tuning process while enabling high generality, efficient configuration search, tuning knowledge reuse, and fast validations.}

\begin{table}[]
    \footnotesize
    \centering
    \setlength{\tabcolsep}{0.2em}
    \caption{Impact of Previous Knowledge on Iteration Rounds and Final Performance}
    \label{tab:workload_performance}

    \begin{tabular}{lccc}
        \toprule
        \multicolumn{1}{c}{\textbf{Workload}} &
        \multicolumn{1}{c}{\textbf{\begin{tabular}[c]{@{}c@{}}With Previous \\ Knowledge\end{tabular}}} &
        \multicolumn{1}{c}{\textbf{\begin{tabular}[c]{@{}c@{}}Without Previous \\ Knowledge\end{tabular}}} &
        \multicolumn{1}{c}{\textbf{\begin{tabular}[c]{@{}c@{}}Achieved Performance \\ (\% of With Knowledge)\end{tabular}}} \\ \midrule
        Fillrandom  & 6  & 15 & 0.84 \\
        Readrandom  & 8  & 12 & 0.82 \\
        MixGraph    & 9  & 17 & 0.76 \\
        \bottomrule
    \end{tabular}
\end{table}

\section{\sysname} \label{sec:design}

In this paper, we propose \sysname, a general, fully automated, end-to-end tuning framework for various storage systems that incorporates LLM reasoning while ensuring both efficiency and robustness. We first discuss the challenges being addressed in \sysname. Then, we present the design and implementation details.
\subsection{Challenges}
\textit{\underline{First}}, the tuning process spans multiple stages, each with distinct contexts, reasoning requirements, and feedback loops. How to decompose the workflow into modular subtasks, define clear scopes and contexts for each, and coordinate reasoning across multiple modules is challenging. \textit{\underline{Second}}, current tuning approaches are tightly coupled to individual storage system design, specific versions, and low-level implementation details, making them brittle and labor-intensive to extend. The key challenge is how to abstract common tuning semantics and design representations that enable LLMs to reason across diverse storage systems without extensive human intervention. \textit{\underline{Third}}, configuration spaces are high-dimensional, and often include complex parameter interactions. Exhaustive search is infeasible, while naive trial-and-error can lead to costly exploration overheads. How to design strategies that help LLMs prioritize promising configurations, prune redundant or low-value trials, and adaptively balance exploration with exploitation?
\textit{\underline{Finally}}, LLM outputs can be hallucinated, and tuning processes may fail due to environmental variability or execution errors. Can we capture, manage, and leverage knowledge from past tuning trials to improve search efficiency and correctness? Also, how to validate LLM-generated suggestions, detect errors, and recover gracefully with minimal human intervention.

\subsection{\sysname Architecture Overview}

The overall architecture of \sysname is shown in Figure~\ref{fig:arc}, \sysname automates workload benchmarking, performance analysis, configuration search, and tuning experiences management with 4 dedicated LLM agents, including Executor, Extractor, Searcher, and Reflector: 
\begin{compactItemize}
\item \textbf{Executor} launches the sandbox engine to deploy the storage system in the targeted setups and run benchmarks based on given configurations and resource constraints. At the same time, it monitors runtime statistics (e.g., CPU utilization, I/O bandwidth) and collects structured (e.g., JSON files) or unstructured (e.g., text logs) output from benchmarking.
\item \textbf{Extractor} analyzes benchmark output and runtime statistics from the Executor, extracts structured performance metrics (referred to as the \textbf{performance digest}), and passes them to the Searcher as benchmarking and deployment results for the given configuration. 
\item \textbf{Searcher} explores the configuration space, proposes and selects the next round of configuration candidates based on the performance digest from the Extractor, and tuning insights (i.e., summary of analysis from past tuning history) from Reflector. It also records tuning experiences (e.g., explored configurations and their performance digest), which are then sent to the Reflector.
\item \textbf{Reflector} collects, analyzes, and manages tuning experiences from the Searcher, and summarizes them into high-level tuning insights and dynamically updates and manages those insights based on feedback from the Searcher.
\end{compactItemize}

\sysname operates through two intertwined iterative cycles during a tuning round, as shown in Figure~\ref{fig:arc}. 1) \textbf{Cycle A} (red lines) involves the \textbf{Executor}, \textbf{Extractor}, and \textbf{Searcher}, which collaboratively propose new configuration candidates, run benchmarks, and analyze the results of each candidate. 2) \textbf{Cycle B} (blue lines) involves the \textbf{Searcher} and \textbf{Reflector}, which collect and transform tuning experiences into high-level insights. Searcher uses these insights to guide the configuration search, and Reflector continuously updates them with new benchmarking results. These two cycles operate continuously, enabling the framework to progressively refine its understanding of storage system behavior and improve storage system tuning efficiency over time.

With dedicated LLM agents for each major task of the tuning process (i.e., Executor, Extractor, Searcher, and Reflector), \sysname decouples the dependencies between storage system-specific knowledge, designs, and executions from the common tuning workflows and handles them via LLM agents. \sysname only requires lightweight interfaces for running benchmarks and adjusting configurations for different storage systems and implementation versions, achieving high generality. Moreover, \sysname provides LLM agents with structured, system-related, task-specific contexts, enabling more accurate reasoning and decision-making at each stage of the tuning process.

\subsection{Automated Benchmarking and Analysis}\label{sec:agent}

In a tuning process, a new configuration is evaluated (e.g., via benchmarks or traces) and compared against other candidate configurations to guide the configuration adjustment. However, these processes are tightly coupled with the storage system and often require custom scripts to automate benchmarking. Furthermore, analyzing the benchmarking results usually requires manual efforts due to the complexity of the data content and format (e.g., runtime statistics, execution logs, system monitoring data, and benchmark outputs). \sysname delegates benchmarking and analysis actions to two specialized LLM-powered agents, \emph{Executor} and \emph{Extractor}.

\begin{figure}
    \centering
    \includegraphics[width=0.99\linewidth]{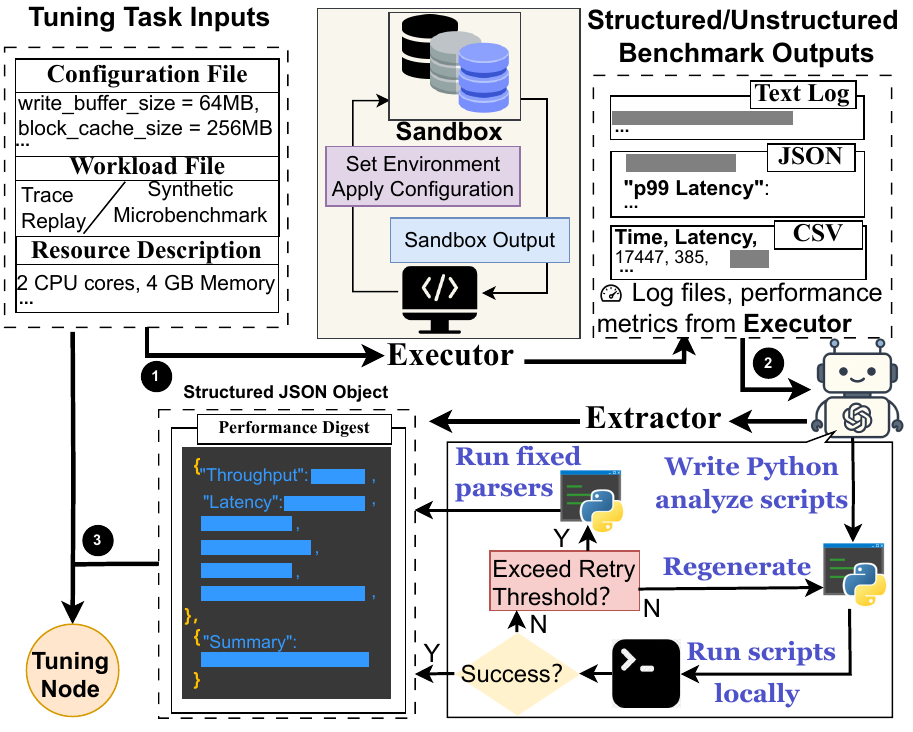}
    \caption{Automated Benchmarking and Analysis 
    } 
    \vspace{-0.5em}
    \label{fig:benchmark}
\end{figure}

\myline{Executor}
As shown in Figure~\ref{fig:benchmark}, the Executor uses a precise specification for each benchmark experiment, including a candidate configuration file (e.g., RocksDB configuration with \textit{write\_buffer\_size = 64MB}, \textit{block\_cache\_size = 256MB}), resource constraints (e.g., 2 CPU cores, 1GB memory), and a specific workload. The workload can be created from synthetic micro-benchmarks (e.g., fillrandom from db\_bench in RocksDB) or trace replay. The Executor sets up the sandbox environment (e.g., Docker), applies the given configuration, and generates the workload within the specified resource envelope to isolate tuning experiments and enable safe, repeatable testing. This design can isolate potential errors or unsafe LLM-generated configurations.
Executor also monitors runtime status (e.g., container health, CPU load, memory usage) and collects all sandbox outputs, including performance metrics from benchmarking output (e.g., throughput and tail latency), storage system operational logs, and any system warnings.

\myline{Extractor}
The massive structured and unstructured output data from Executor raises the challenge of how to analyze them for configuration exploration. Conventional approaches rely on manual parsing or rigid, fixed-format parsers, which are labor-intensive and difficult to scale \cite{cao_characterizing_2020}. When output formats change with new system releases, these methods require costly re-engineering. While LLMs can directly analyze such files to extract performance metrics, directly feeding massive logs into an LLM risks token overflows and hallucinations.

Therefore, \sysname exploits the LLM’s programming capabilities and uses a separate LLM agent, Extractor, to generate a targeted Python parser on demand. We provide the LLM with descriptions of the performance metrics of interest (e.g., definitions of throughput and latency), together with relevant system information (e.g., system version), and data samples (e.g., log snippets). Guided by this context, the LLM infers the possible output formats and synthesizes Python code to process them (e.g., import different libraries to handle text, JSON, or CSV files). Then, the Extractor executes the Python script locally to transform raw outputs into structured key–value pairs to store extracted performance metrics and non-metric information. This design enables \sysname to adapt on the fly to the unique logging formats of various storage systems without heavy engineering effort. Finally, Extractor constructs the \emph{Performance Digest}, a structured JSON object that includes quantitative performance metrics and qualitative evaluation summary (i.e., a concise interpretation of benchmark results, including key trends, trade-offs, or any anomalies generated by LLM). The performance digest, combined with the configuration file, workload description, and system resource information, forms the tuning node in the search tree used by Searcher for configuration exploration (detailed in Section \ref{sec:detail}).

\myline{Validation} To ensure the reliability of the generated Python scripts, Extractor incorporates a self-correcting mechanism. As shown in Figure~\ref{fig:benchmark}, if a generated script fails to execute or produces invalid output (e.g., incorrect types), this failure is detected by a pre-defined checker and reported back to the LLM to generate a refined script. If regeneration attempts exceed a predefined threshold, the system rolls back to pre-defined, fixed-format human-written parsers. A script may run successfully but still compute metrics incorrectly. To address this, Extractor performs pre-defined abnormal value checks (e.g., throughput exceeding hardware limits) to identify suspicious outputs and trigger regeneration.

\subsection{Insight-Driven Configuration Exploration}\label{sec:detail}

While LLM can directly propose candidate configurations based on the performance digest from the last tuning cycle, as Elmo-Tune did \cite{Thakkar2024can}, it cannot efficiently explore configurations that need past tuning experiences. The LLM reasoning is session-bound and limited by context windows, making it impractical to feed all past trial results without performance degradation or hallucinations \cite{li2024alr}. In addition, LLMs lack iterative feedback integration and tend to generate stochastic, unguided exploration paths, which often lead to redundancy and incomplete coverage \cite{kostikova2025lllms}.

We propose to integrate LLM with a tree search process for configuration exploration, which addresses these limitations by structuring exploration into explicit steps, recording past outcomes, and guiding the search toward promising directions. By maintaining parent–child relationships between configuration candidates, tree search expands promising branches while pruning unpromising ones. This structure combines planning, branching, and feedback-driven prioritization, enabling LLMs to concentrate their reasoning on high-potential areas of the configuration space.

\sysname achieves this by delegating the configuration tree search task to a specialized LLM agent, the Searcher. Unlike traditional tree search, which depends on scalar rewards or hand-tuned utility functions (demanding substantial human effort to predefine and tune), the Searcher leverages high-level tuning insights, which are summarized by the LLM from past tuning trials (See details in Section~\ref{sec:insights}), to decide how and where to explore.

\mynewline{How to Propose Configuration Candidates?}
The exploration process begins with the initial configuration (e.g., the default configuration), which serves as the root node in the search tree. Tuning insights, expressed in natural language, capture plausible relationships between configuration changes and observed storage system behavior. For instance, an insight might state that increasing the write buffer pool size significantly improves throughput for RocksDB. Guided by these insights, the Searcher proposes a set of configuration candidates, which become the child nodes of the current node (the root in this case) in the search tree. The LLM is directed to prioritize exploration consistent with these insights, such as re-balancing the write buffer and block cache size.

\begin{figure}
    \centering
    \includegraphics[width=1\linewidth]{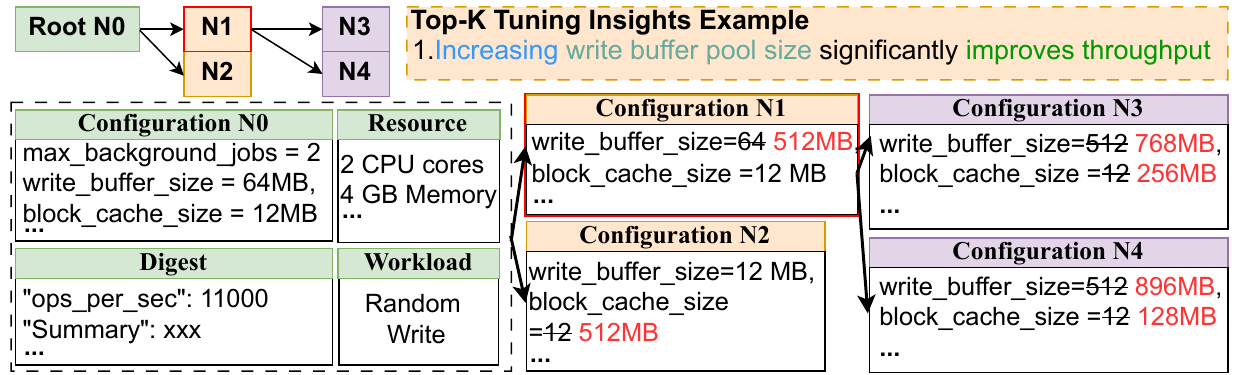}
    \caption{Insight-Driven Configuration Exploration
    } 
    \vspace{-0.5em}
    \label{fig:tree}
\end{figure}

\begin{figure}
    \centering
    \includegraphics[width=0.99\linewidth]{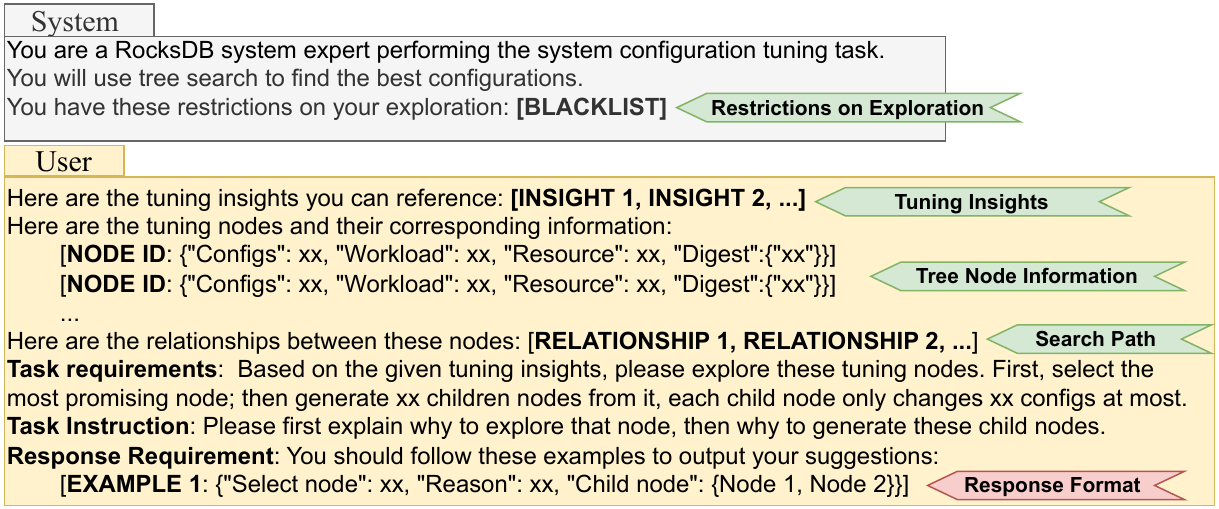}
    \caption{Example Prompt for Configuration Proposal}
    \vspace{-0.5em}
    \label{fig:Prompt_1}
\end{figure}

As shown in Figure~\ref{fig:tree}, consider tuning RocksDB under a random write workload with 2 CPU cores and 1 GB of memory, where the initial configuration (root node) sets \textit{max\_background\_jobs=2}, \textit{write\_buffer\_size=64MB}, and \textit{block\_cache\_size=12MB}. First, Searcher constructs a structured prompt that incorporates the current tuning node (root node), task descriptions, and tuning insights. Figure~\ref{fig:Prompt_1} shows how the prompt template organizes these elements into a natural-language instruction. This prompt is then passed to the LLM to propose child configurations for exploration. Then, the Searcher expands the root node by proposing two candidate configurations, increasing \textit{write\_buffer\_size} and \textit{block\_cache\_size} to 512MB each to fully utilize the 1 GB memory budget. Next, the Searcher explores re-balancing strategies within the same memory budget. Guided by the insight that increasing \textit{write\_buffer\_size} can improve throughput under write-intensive workloads, it reallocates memory by assigning 768MB to \textit{write\_buffer\_size} and 256MB to \textit{block\_cache\_size}. In practice, Searcher may propose multiple child nodes.

\mynewline{Which Configuration Should Be Further Explored?}
After the Searcher proposes candidate configurations, a new question arises: which child node should be explored in the next round? Before selecting the most promising node, we send the benchmark tasks, including candidate configurations, resource descriptions, and workload descriptions, to the Executor. The Executor benchmarks these candidates, and the Extractor analyzes the results to generate a new set of performance digests. This performance digest serves as the utility for each node. As shown in Figure~\ref{fig:Prompt_1}, we combine the performance digests with the corresponding tuning nodes, insert them into the prompt template, and guide the LLM to compare these nodes, reason about which node is more promising, and select it for the next round of exploration.

Since the performance digest is a structured summary rather than a directly comparable numeric value, the LLM relies on its reasoning capabilities to weigh trade-offs and select the most promising node for exploration. This process may choose a suboptimal node, for example, a configuration with the second-highest throughput. We consider this acceptable because even if the current node does not achieve the highest performance, exploring it can still provide valuable information. Furthermore, if future iterations reveal poor outcomes for this node, the LLM can return to select the optimal node in subsequent rounds. In addition, we can also add specific insights to guide the LLM toward prioritizing the optimal node. For instance, an insight stating that “throughput is our main target” encourages the selection of higher-throughput configurations. This soft comparison approach enables the LLM to explore promising directions while maintaining a degree of randomness in its exploration.

\mynewline{How to Validate Configuration Candidates?} Given the inherent uncertainty of LLM reasoning, \sysname relies on a two-layer validation strategy to ensure the correctness and safety. In the first layer, users can specify domain constraints, such as persistence guarantees or concurrency limits. The LLM applies these constraints to filter out unsafe or irrelevant candidate configurations before evaluation. Configurations that pass this filtering stage undergo empirical validation in the second layer. Here, the Extractor performs consistency checks using pre-defined scripts. These checks include detecting anomalous metrics, enforcing blacklists to reject unsafe configuration changes, and reporting errors back to the LLM. This layered validation ensures that while LLMs provide flexible semantic reasoning, all tuning decisions remain firmly grounded in reliable, observable feedback.

\mynewline{When to Terminate the Exploration?} To ensure bounded search, \sysname supports multiple termination strategies. The search process halts automatically when the performance improvement between iterations falls below a threshold (e.g., 1\% over three rounds), indicating convergence. Alternatively, searching can be terminated based on user-specified constraints such as maximum LLM token usage, total tuning time, or a system resource budget. 
These controls provide both flexibility and safety, allowing \sysname to be deployed in diverse operational settings ranging from offline experimentation to online tuning in production environments.

\begin{figure*}
    \centering
    \includegraphics[width=1\linewidth]{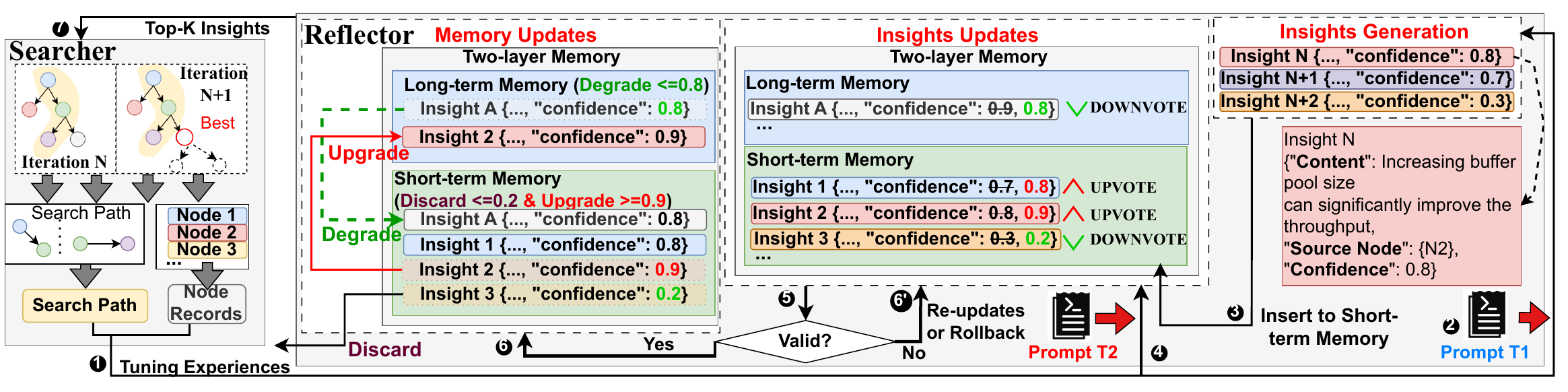}
    \caption{Tuning Insight Generation and Management. The LLM updates an insight’s confidence through \textbf{Upvote} or \textbf{Downvote} based on observed results (e.g., Node 2 shows performance consistent with Insight N’s prediction, while Insight 3 contradicts it). Insights with high confidence are promoted from STM to LTM as validated, reusable knowledge, while low-confidence insights are discarded or demoted for further evaluation.
    } 
    \vspace{-0.3em}
    \label{fig:reflector}
\end{figure*}

\subsection{Tuning Insights Management}\label{sec:insights}

Tuning insights play a critical role in configuration exploration. The Searcher relies on them to decide how and where to explore within the complex configuration space.
\begin{figure}
    \centering
    \includegraphics[width=0.99\linewidth]{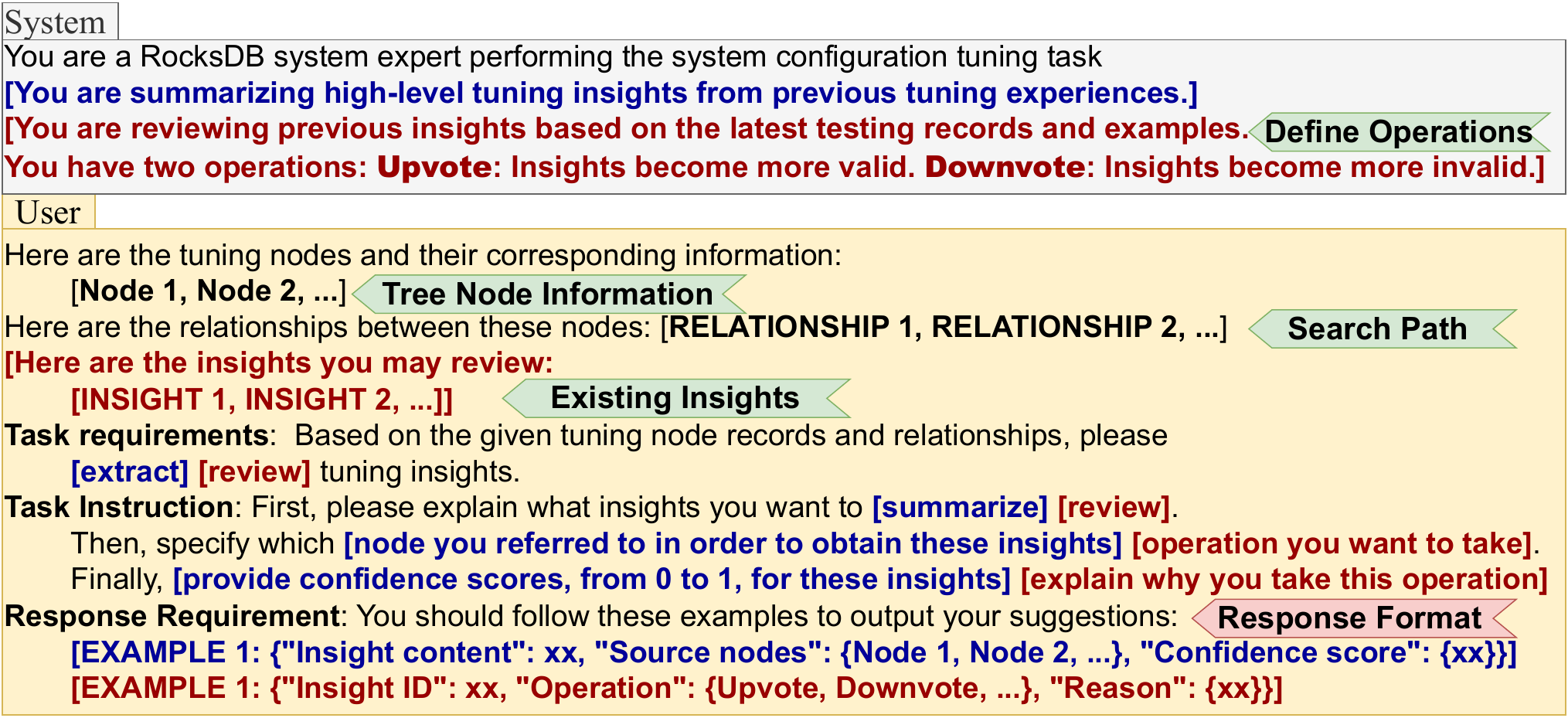}
    \caption{Prompt for Insight Generation and Updating}
    \vspace{-0.5em}
    \label{fig:Prompt_2}
\end{figure}

\myline{Insights Generation}
As shown in Figure~\ref{fig:reflector}, insights are derived from past tuning experiences. In each search iteration, the Searcher collects tuning experiences, which include node records (e.g., executed configurations and their performance digests) and the search paths (e.g., the relationships between explored nodes), and sends them to the Reflector. Those records are inserted into a prompt (as shown in Figure~\ref{fig:Prompt_2} (Blue Text segments)). Reflector is then instructed to compare performance digests across different nodes, learn the relationship within these nodes, and infer generalized rules, such as which parameter adjustments consistently lead to performance gains or losses.
Each insight is expressed in natural language and is associated with an initial confidence score, which is assigned by the Reflector based on its initial assessment and updated as more evidence becomes available. The confidence score denotes the confidence level of the insight, reflecting how strongly it is supported by past evidence. Besides, each insight also records the source nodes from which it was inferred.

\myline{Insights Management} After generating tuning insights, managing and selectively providing them becomes critical for improving configuration exploration efficiency. We may use Retrieval-Augmented Generation (RAG) to manage the insights \cite{lewis2020retrieval}, which retrieves relevant insights via vector similarity and injects them as auxiliary context into the LLM’s prompt. However, RAG is insufficient for the storage system tuning task. First, it relies on a static, homogeneous knowledge base, making no distinction between volatile, session-specific insights and validated, reusable ones. Second, it retrieves knowledge solely by embedding similarity, which is brittle when insight utility depends on subtle workload dynamics. Embedding similarity only captures surface-level context similarity and lacks a feedback loop, preventing real tuning outcomes from reinforcing or revising stored insights. 

To address those limitations, \sysname adopts two key mechanisms in Reflector. First, it maintains a layered memory structure: a Short-Term Memory (\textbf{STM}) for session-specific insights and a Long-Term Memory (\textbf{LTM}) for validated, reusable insights across sessions. Second, it applies a \emph{Confidence Scoring-based Management}, where retrieval considers both contextual similarity and dynamically updated confidence scores that reflect empirical tuning outcomes. STM serves as a workspace for newly generated or session-specific insights, while LTM accumulates validated insights that have demonstrated consistent effectiveness across different tuning executions. Each new insight is first placed into STM with a provisional confidence score reflecting the LLM’s initial assessment of its potential utility. Since such confidence scores are unreliable at generation time, STM functions as a staging area where tentative knowledge is empirically tested against benchmark results. 

For example, under a write-heavy workload on a 2-core system for RocksDB, an STM insight might be “increasing \textit{max\_background\_jobs} from 2 to 4 could improve compaction throughput.” Such an insight is specific to the current session and may not generalize. As evidence accumulates, insights in STM are either discarded if contradicted or promoted to LTM once repeatedly validated. LTM, in turn, stores cross-session knowledge such as “larger \textit{block\_cache\_size} consistently reduces read latency”. This separation of transient and persistent knowledge enables \sysname to adapt dynamically to workload-specific behaviors while gradually consolidating durable expertise. 

The Reflector leverages the Confidence Scoring-based Management to adjust and manage the insights. The confidence score of an insight represents the level of confidence, indicating how strongly an insight is expected to produce the predicted effect. Each insight receives an initial confidence score from the LLM during generation, but this score is provisional and continuously updated as new benchmarking evidence arrives. We adjust the score through two LLM-driven operations: \textbf{Upvote}, which raises confidence when observed benchmarking outcomes align with the insight, and \textbf{Downvote}, which lowers confidence when results contradict the guidance. As shown in Figure~\ref{fig:Prompt_2} (Red Text segments), the Reflector applies dedicated prompts to interpret benchmark results and update the confidence scores of insights. Over time, this dynamic updating allows the Reflector to validate, demote, or discard insights, ensuring that only empirically grounded knowledge persists in LTM.

\myline{Insights Retrieval}
The Reflector combines each insight’s updated confidence score with its contextual similarity (e.g., cosine similarity \cite{lewis2020retrieval}) to produce a similarity metric weighted by confidence. The top-$k$ insights (e.g., $k=10$) according to this metric are then returned to the Searcher, guiding the next round of configuration proposals with both empirically validated and contextually relevant knowledge. In this way, the retrieval workflow closes the loop between reasoning, empirical validation, and insight management, ensuring that subsequent search steps are progressively better informed and increasingly effective.

\myline{Validation} 
\sysname includes a lightweight Validator that cross-checks LLM-proposed confidence updates against actual benchmarking results. For each insight, the Validator verifies whether associated nodes show performance changes consistent with the insight's predicted effect. Confidence updates are then accepted or rejected based on this verification, ensuring that insights reflect concrete benchmark outcomes rather than relying solely on LLM reasoning.

\section{Evaluations}
\subsection{Prototype Implementation}
We implement \sysname in Python (v3.10) as a modular, LLM-driven framework composed of collaborative LLM agents. 
Our evaluation spans three representative storage systems: 1) the most widely used key-value store RocksDB (v8.8.1) \cite{rocksdb}, 2) a powerful and modularized cache engine CacheLib developed by Meta \cite{cachelibcode}, and MySQL InnoDB \cite{innodb_mysql}, to cover a diverse range of storage systems. Note that for MySQL we focus on the configuraitons of its default storage engine InnoDB, and leave other optimizations (e.g., query-level tuning) as future work.

We extend standard benchmarking tools, including YCSB for RocksDB \cite{dbbench},
CacheBench for CacheLib \cite{cachebench}, and TPC-H/C workloads for MySQL \cite{tpc-h}, to support workload generation and configurable tuning parameters. By default, \sysname uses OpenAI’s GPT-4o \cite{gpt_4o_welcome_page} as its backend LLM model, and the model can be replaced. The source code for \sysname is available on GitHub \cite{gdaythcli_gdaythcliomnitune_2025}.

\subsection{Experimental Setup}

\textbf{Hardware and System Setup.} 
Our experimental evaluation is conducted on a workstation equipped with Intel(R) Core(TM) i9-10920X CPU @ 3.50GHz, 32 GB of RAM, a 512GB NVMe SSD running Ubuntu 24.04.1 LTS. We leverage cgroups \cite{cgroup} and Docker \cite{noauthor_docker_nodate} to isolate workloads and enforce resource constraints. Unless otherwise specified, all tests are limited to 2 CPU cores, 4GB of DRAM, and 4GB of swap space.

\smallskip
\noindent\textbf{Baselines.}
We use the default configuration file provided by each application as our primary baselines, including the default RocksDB configurations (\textbf{\textit{RocksDB-Default}}), the default configurations of CacheLib (\textbf{\textit{CacheLib-Default}}), and the default configurations of InnoDB (\textbf{\textit{InnoDB-Default}}). In addition, we compare \sysname against state-of-the-art tuning solutions. Specifically:
\begin{compactItemize}
\item \textbf{RocksDB}: \textbf{\textit{ADOC}}\cite{yu2023adoc} (heuristic-based tuner), \textbf{\textit{Endure}}\cite{Endure} (ML-based tuner), \textbf{\textit{ELMo-Tune}}\cite{Thakkar2024can} (LLM-based state-of-the-art tuner), \textbf{\textit{LLM-Default}} (simple one-shot LLM tuning), and \textbf{\textit{\sysname}} (our solution).

\item \textbf{CacheLib}: \textbf{\textit{LLM-Default}} and \textbf{\textit{\sysname}}.

\item \textbf{InnoDB}: \textbf{\textit{SMAC}}\cite{JMLR:v23:21-0888_smac} (Bayesian optimization for tuning), \textbf{\textit{DDPG}}\cite{fakoor2020ddpgstrivingsimplicitycontinuouscontrol_ddpg} (Reinforcement-learning based tuning), \textbf{\textit{$\lambda$-Tune}}\cite{giannankouris2024lambda, giannakouris2024demonstrating} (a LLM-based tuner), \textbf{\textit{LLM-Default}}, and \textbf{\textit{\sysname}}. 

\end{compactItemize}

\begin{figure*}
    \centering
    \includegraphics[width=0.99\linewidth]{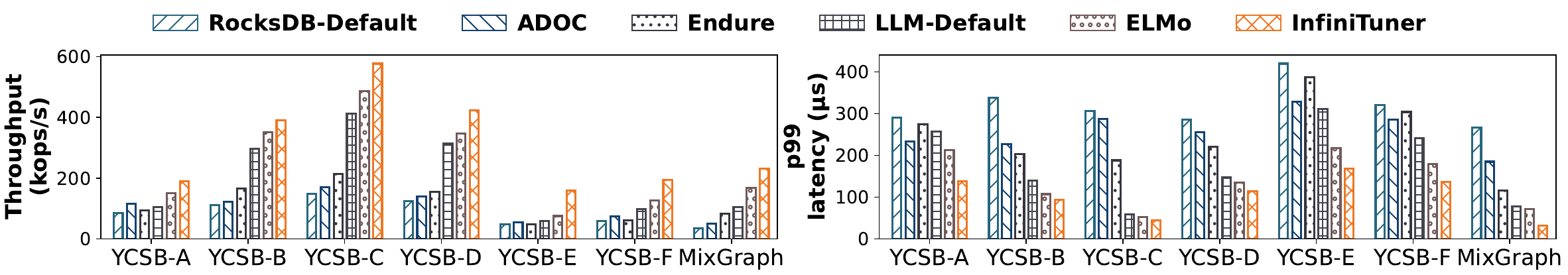}
    \caption{Performance Comparison under YCSB and Mixgraph Workload in RocksDB}
    \label{fig:macro-benchmarks-ycsb}
\end{figure*}

\smallskip
\noindent\textbf{Workloads and Benchmarks.}
We evaluate \sysname across a variety of real-world workloads tailored to each system under test. For RocksDB, we use macro-benchmarks based on Yahoo's YCSB \cite{ycsb} and Meta's MixGraph \cite{cao2020characterizing} to capture realistic access patterns. Unless otherwise specified, all write-intensive workloads operate on 50 million key-value (KV) pairs, and all read-related tests access at least 20 million KV pairs, with each key and value set to 16 bytes and 48 bytes, respectively. For CacheLib, we evaluate three workload profiles: write-intensive, read-intensive, and mixed workloads. Each test involves a total of 50 million operations, the average cache item size is set to 128 bytes, and the access pattern is defined by a configurable lookup-to-insert ratio, set to 9:1 for read-intensive workloads, 1:9 for write-intensive workloads, and 5:5 for mixed workloads. For MySQL InnoDB, we benchmark the system using TPC-C and TPC-H workloads\cite{tpc-h, noauthor_tpc-c_nodate}, with database sizes set to 1GB and 5GB, respectively. All reported results are the average of three or more runs.

\noindent\textbf{Performance Metrics.}
Since LLM-driven tuning is a fully new research area, traditional measurement metrics like eventual throughput or latency improvement are insufficient to capture the unique dynamics of LLM-based solutions. LLMs introduce reasoning overheads and semantic decisions that can yield both significant gains and unpredictable outcomes. Therefore, we propose \textbf{four new measurement metrics} tailored for LLM-driven tuning, which can better guide the designs and evaluations.
\textbf{Max Performance Gain (MPG)} captures the relative improvement over the baseline performance metrics (e.g., throughput). \textbf{Token Cost to 95\% Max Performance (TC95)} measures the total number of LLM tokens consumed to reach 95\% of the peak performance. We use 95\% of peak performance instead of the absolute maximum because it better reflects typical token usage in practice. \textbf{Token Efficiency (TE)} quantifies how much performance gain is achieved per 1,000 tokens used, helping assess the cost-effectiveness of tuning. \textbf{Token-Weighted Error Rate (TWER)} represents the number of invalid or error-prone configurations generated per 1,000 tokens. Invalid configurations include syntax errors, unrecognized options, or configurations that result in execution failures. These metrics collectively offer insight into the quality, efficiency, and robustness of the LLM-driven tuning solutions.

\begin{figure}
    \centering
    \includegraphics[width=\linewidth]{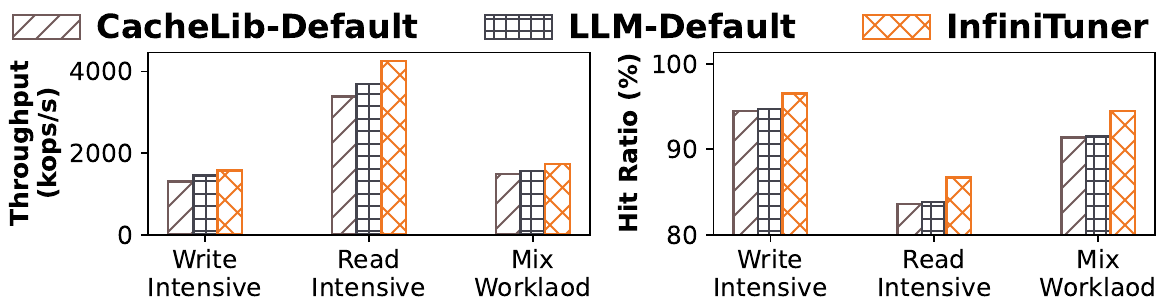}
    \caption{Performance with Different Workload in CacheLib}
    \vspace{-0.5em}
    \label{fig:macro-benchmarks-cache}
\end{figure}

\begin{figure}
    \centering
    \includegraphics[width=\linewidth]{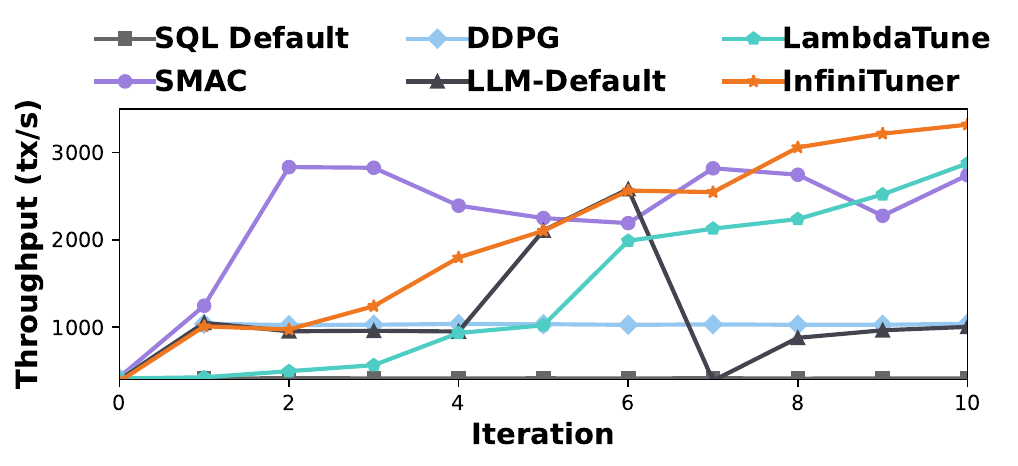}
    \caption{Best Performance over Iterations on InnoDB. }
    \vspace{-0.5em}
    \label{fig:macro-benchmarks-sql}
\end{figure}

\subsection{Overall Performance Evaluation} 
\myline{Key-Value Stores} We assess \sysname’s performance on real-world benchmarks using MixGraph \cite{cao2020characterizing} and YCSB \cite{ycsb} workloads for RocksDB. With MixGraph workload, as shown in Figure~\ref{fig:macro-benchmarks-ycsb}, \sysname achieves up to a 575\% increase in throughput and a reduction in latency by up to 88\% compared to other baselines. Under the YCSB workload, \sysname achieves throughput improvements of up to 111\% and p99 latency reductions of up to 56\% compared to ELMo-Tune, the current state-of-the-art LLM-based tuning framework for RocksDB. These results demonstrate \sysname’s more effective use of LLMs and structured feedback. ADOC, Endure, rule-based, and ML-based tuning solutions improve throughput over the RocksDB-Default baseline by up to 43\% and 141\%, respectively. However, they still fall short of \sysname: \sysname surpasses these two methods by up to 369\% in throughput. Additionally, compared to the LLM-Default baseline, a naive one-shot interaction with an LLM lacking insight feedback and validation, \sysname achieves a throughput improvement of up to 80\% and significantly more stable tail latencies. These results highlight the importance of structured tuning experiences, insight-driven search, and feedback validation in enabling efficient and reliable LLM-powered tuning.

\myline{Cache Systems} For CacheLib, we evaluate \sysname across three workload types: write-intensive, read-intensive, and mixed. We use the cachebench tool with configurable access patterns and object sizes to simulate realistic cache usage scenarios. To the best of our knowledge, no existing LLM-based tuning framework has been applied to CacheLib, making \sysname the first to address this configuration space using an LLM-driven approach. As shown in Figure~\ref{fig:macro-benchmarks-cache}, compared to the CacheLib-Default baseline, \sysname achieves up to a 26\% improvement in throughput and improves cache hit ratio by up to 3.1 percentage points across workloads. Specifically, for the write-intensive workload, \sysname increases the hit ratio from 94.5\% to 96.5\%, and for the read-intensive workload, the hit ratio improves by 2.9 percentage points, demonstrating more efficient memory utilization and improved eviction policy decisions. Additionally, it outperforms the LLM-Default baseline by up to 15\%, illustrating the advantage of \sysname’s insight-driven, feedback-aware tuning in optimizing complex caching systems.

\myline{SQL Storage Engines} For InnoDB, the default storage engine in MySQL, \sysname is evaluated using two widely adopted transactional and analytical benchmarks: TPC-C and TPC-H, respectively. 
\sysname consistently outperforms all baselines across both benchmarks. As shown in Figure~\ref{fig:macro-benchmarks-sql}, under the TPC-C workload, \sysname increases throughput by up to 709\% over InnoDB-Default and 29\% over LLM-Default. \sysname achieves over 15\% higher throughput compared to $\lambda$-Tune. In the TPC-H workload, \sysname reduces query latency by up to 71\% (from 8.9s to 2.5s) across all baselines, showcasing its ability to improve analytical query performance through better buffer pool sizing and I/O tuning. This underscores the effectiveness of its insight-driven and feedback validation tuning, even in relational database storage engines.

\newcolumntype{Y}{>{\centering\arraybackslash}X}

\newcolumntype{L}[1]{>{\raggedright\arraybackslash}m{#1}}

\begin{table}[t]
  \centering
  \footnotesize
  \setlength{\tabcolsep}{2pt}
  \caption{Standardized Metric Analysis.}
  \label{table:comparison-metrics}
  \begin{tabularx}{\linewidth}{L{3.8cm}YYY}
    \toprule
    \textbf{Metric} & \textbf{\sysname} & \textbf{ELMo-Tune} & \textbf{LLM-Default} \\
    \midrule
    Max Performance Gain          & 5.75  & 3.88  & 2.04  \\
    Token Cost to 95\% Max        & 56K   & 138K  & 214K  \\
    Token Efficiency              & 0.10  & 0.02  & 0.009 \\
    Token-Weighted Error Rate     & 0.017 & 0.024 & 0.028 \\
    \bottomrule
  \end{tabularx}
\end{table}

\myline{Standardized Metric Analysis}
Table~\ref{table:comparison-metrics} highlights the advantages of \sysname over the LLM-Default baseline across several standardized evaluation dimensions. First, \sysname achieves a significantly higher peak performance, demonstrating its ability to discover more effective configurations. More importantly, it reaches 95\% of its peak performance using fewer LLM tokens (56K vs. 214K), which translates to both lower latency in interaction rounds and reduced monetary cost. This improvement in cost-efficiency is further underscored by its higher token efficiency, achieving 11× and 5× improvements compared to LLM-Default and ELMo-Tune, respectively.  Beyond efficiency, \sysname exhibits greater robustness. It accumulates fewer errors in the tuning process, as evidenced by a lower Token-Weighted Error Rate (TWER = 0.017 vs. 0.028), indicating better stability and reduced need for human correction or fallback mechanisms. 

\begin{figure}[t]
    \centering
    \includegraphics[width=0.99\linewidth]{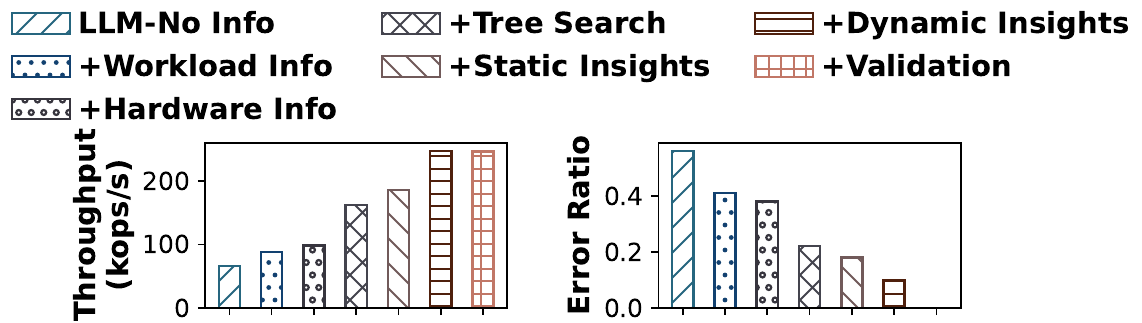}
    \caption{Ablation Study}
    \vspace{-0.5em}
    \label{fig:ablation}
\end{figure}

\begin{table}[ht]
\centering
\caption{Performance comparison across different versions on the \textit{fillrandom} workload (throughput in kops/s). \xmark~indicates the method fails to run.}
\small
\begin{tabular}{lcccccc}
\toprule
 & \multicolumn{4}{c}{RocksDB} & \multicolumn{2}{c}{LevelDB} \\
\cmidrule(lr){2-5} \cmidrule(lr){6-7}
Method & 5.7.1 & 6.11.6 & 7.5.3 & 8.8.1 & 1.23 & 1.21 \\
\midrule
Default     & 47.6  & 126 & 196 & 234 & 108 & 64  \\
SILK        & 114 & \xmark   & \xmark   & \xmark   & \xmark   & \xmark   \\
ELMO        & \xmark    & \xmark   & 210 & 242 & 119 & 72  \\
\sysname & 126  & 142 & 221 & 246 & 124 & 74  \\
\bottomrule
\end{tabular}
\label{tab:dbms-versions}
\end{table}

\subsection{Ablation Study}
To understand the impact of progressively enriching the tuning context for LLM-driven configuration generation, we conduct an ablation study on RocksDB using MixGraph. In addition to varying the levels of input information, we also evaluate the contributions of our design components, including tree search, dynamic tuning insights, and the validation checker. We measure how each element affects both throughput and error ratio to quantify their individual and combined benefits.

The evaluation is conducted with the following setups:

\textbf{LLM-No Info}: The baseline where the LLM receives only a general tuning task description prompt, with no other tuning context.
\textbf{+Workload Info}: Adds workload benchmarking results.
\textbf{+Hardware Info}: Adds hardware metadata, such as CPU and memory information.
\textbf{+Tree Search}: Adds Tree Search without insights.
\textbf{+Static Insights}: Adds static insights.
\textbf{+Dynamic Insights}: Adds dynamic insights (insights management).
\textbf{+Validation}: Combines validation mechanisms.

As shown in Figure~\ref{fig:ablation}, each additional layer of context contributes to measurable improvements. Starting with the LLM without any guidance, throughput is the lowest (65,908 ops/sec) with a high error ratio (56\%). Adding \textbf{workload information} increases throughput to 87,664 ops/sec and reduces the error ratio to 41\%. Incorporating \textbf{hardware information} further improves throughput to 98,470 ops/sec and lowers the error ratio to 38\%. Introducing tree search context boosts throughput to 161,847 ops/sec and reduces the error ratio to 22\%. Adding \textbf{static insights} from historical tuning records further increases throughput to 185,542 ops/sec and lowers error ratio to 18\%. Incorporating \textbf{dynamic insights}, which include runtime feedback and performance refinement, raises throughput to 247,257 ops/sec and reduces the error ratio to 10\%. Finally, the full \sysname setup with \textbf{validation} mechanisms achieves 246,352 ops/sec and eliminates all invalid configuration attempts.

\subsection{Sensitivity Analysis}
To evaluate the robustness of \sysname, we conduct a sensitivity analysis across a variety of resource and model-level parameters. This study demonstrates how the performance of \sysname responds to changes in system version diversity, exploration child node, number of selected top insights, and the capabilities of the underlying LLM.

\myline{Different Storage System Implementations and Version}
Table~\ref{tab:dbms-versions} reports throughput under the \textit{fillrandom} workload of RocksDB and LevelDB (two different implementations of LSM-based key-value stores) across multiple versions. We observe that \sysname consistently achieves the highest performance and runs across all tested versions, while Elmo-Tune and SILK\cite{balmau2019silk} (an I/O scheduler for RocksDB) fail on several versions due to compatibility issues.
This highlights \sysname’s ability to deliver robust and generalizable performance across diverse system versions.

\begin{table}[t]
    \footnotesize
    \centering
    \caption{Impact of Different Number of Child Nodes.}
    \label{tab:child-nodes}
    \begin{tabular}{cccccc}
        \hline
        \textbf{Child Nodes} & \textbf{1} & \textbf{2} & \textbf{3} & \textbf{4} & \textbf{5} \\
        \hline
        Max Performance Gain & 2.15 & 3.58 & 5.75 & 5.86 & 5.92 \\
        Token Efficiency     & 0.07 & 0.09 & 0.10 & 0.07 & 0.06 \\
        \hline
    \end{tabular}
\end{table}

\myline{Different Number of Child Node}
Table~\ref{tab:child-nodes} shows the impact of varying the number of child nodes on tuning effectiveness. Increasing the number of child nodes generally improves maximum performance gain, with diminishing returns beyond three nodes. However, token efficiency does not scale proportionally and even decreases when the branching factor becomes too large. This highlights a trade-off between performance gain and efficiency, suggesting that moderate branching achieves the best balance.

\begin{table}[t]
    \footnotesize
    \centering
    \caption{Impact of Different Numbers of Insights.}
    \label{tab:topk-insights}
    \begin{tabular}{cccccc}
        \hline
        \textbf{K Value} & \textbf{1} & \textbf{2} & \textbf{4} & \textbf{8} & \textbf{16} \\
        \hline
        Max Performance Gain & 1.48 & 2.49 & 4.68 & 5.84 & 5.89 \\
        Token Efficiency     & 0.04 & 0.06 & 0.09 & 0.11 & 0.04 \\
        \hline
    \end{tabular}
\end{table}

\myline{Different Number of Insight}
As shown in Table~\ref{tab:topk-insights} compares the effect of different numbers of Top-K insights on tuning performance. As K increases from 1 to 8, the maximum performance gain improves significantly, indicating that leveraging more insights helps the tuner explore better configurations. However, increasing K beyond 8 yields diminishing returns, and token efficiency drops for K = 16 due to higher token overhead during reasoning. This demonstrates that choosing an appropriate number of Top-K insights is important to balance performance gains and efficiency.

\begin{figure}[t]
    \centering
    \includegraphics[width=\linewidth]{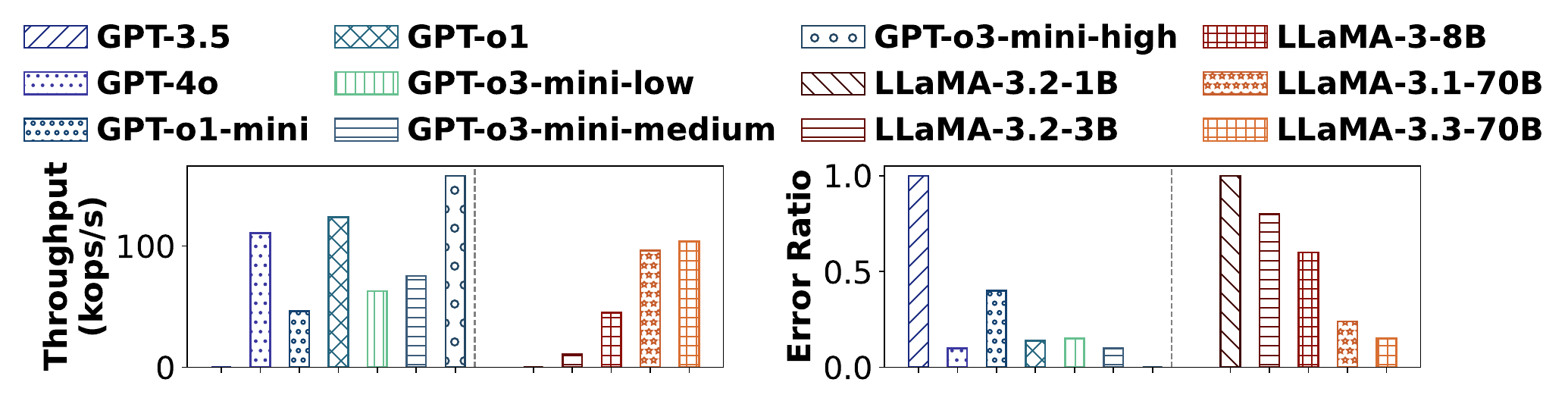}
    \caption{\sysname on Different LLM Models}
    \label{fig:Different_LLM_Models}
\end{figure}

\myline{LLM Model Selection.} As shown in Figure~\ref{fig:Different_LLM_Models}, we evaluate \sysname with a range of LLMs to understand the impact of model size and reasoning capabilities on tuning quality. Specifically, we utilize OpenAI’s GPT-3.5, GPT-4o, o1, o1-mini, and o3-mini models. Where applicable, we also analyze the effect of different reasoning levels within the o3-mini family. To explore the trade-offs between small and large open-source models, we select various configurations from the LLaMA series—namely, LLaMA-3.2-1B, 3.2-3B, 3.1-8B, 3.1-70B, and 3.3-70B—all of which publicly disclose model sizes. The results show that larger, reasoning-capable models consistently yield higher-quality insights and tuning decisions. In contrast, models with limited reasoning ability—such as o1-mini and the low- and medium-capacity variants of o3-mini—perform worse than GPT-4o, which, while powerful, lacks explicit reasoning prompts. 
These results suggest that even a high-capacity model without dedicated reasoning prompts can achieve strong performance, but enabling robust reasoning capabilities further improves tuning effectiveness. Overall, these results show that \sysname can work with different models effectively, and it is expected that the tuning effectiveness will scale as  LLM evolves

\section{Lessons Learned}\label{sec:key-findings}

In this section, we summarize a few key lessons learned from building and evaluating \textit{\sysname}:

\myline{A small set of key insights dominates the search trajectory} 
\noindent
The search process is mainly guided by a small set of human-auditable high-level insights. Treating these insights as first-class artifacts to guide exploration significantly reduces invalid attempts and costs. Critically, these general insights are transferable across heterogeneous systems, effectively accelerating cold starts.

\myline{Proposal of new configurations is the primary error source}
\noindent
Errors primarily arise not from editing existing configurations, but from proposing changes to \emph{untouched} configurations, which trigger string/enum mismatches and version incompatibilities. Systematic numerical mistakes (e.g., unit confusion, budget violations) are also common. These are best mitigated by pre-execution schema validation and enforcing normalized budget caps, rather than relying on reactive fixes.

\myline{Closed-loop agent architecture outperforms monolithic model reasoning}
\noindent
A closed-loop, multi-agent architecture with clear role separation delivers more robust and stable gains than increasing the underlying model's size. This design is largely autonomous for well-documented open-source systems but benefits from lightweight retrieval of local documentation to stabilize performance on closed-source or rapidly evolving ones.

\myline{Human-in-the-loop shifts from low-level tuning to high-level strategy}
\noindent
With \sysname, the human's role evolves from manual tuning to strategic oversight. By providing explicit guardrails (e.g., budgets, SLOs), experts enable the agent to explore safely and broadly. Their primary value thus becomes curating high-impact insights and adjudicating trade-offs across non-aligned objectives, which yields auditable improvements with minimal manual intervention.

\section{Related Work}
\noindent
{\bf Analyzing Software Configurations.} Great efforts exist to analyze and/or test configurations of various software systems~\cite{cdep-fse20,ctest-osdi20,conferr-dsn08,spex-sosp13,cao2017performance-fast17,cao2018towards,carver-fast20,ConfigEverywhere-ICSE14,ArielRabkin-ICSE11,confd-fast23}. For example, SPEX~\cite{spex-sosp13} analyzes configuration constraints and uses inferred constraints to harden systems against misconfigurations.
Similarly, cDEP~\cite{cdep-fse20} and ConfD~\cite{confd-fast23} analyze configuration dependencies in cloud systems (e.g., Hadoop and OpenStack) and file systems (e.g., Ext4, XFS), respectively. 
Insights derived from these efforts could potentially be leveraged to guide auto-tuning and are complementary to \sysname.

    \myline{Heuristic-based Approaches} To alleviate tedious manual configuration, 
    heuristic approaches~\cite{tang2021learning, carver-fast20} encode expert knowledge into predefined rules or strategies to guide parameter adjustments. For example, a workload-aware heuristic can recommend a larger cache size for read-heavy workloads or an infrequent compaction policy for write-heavy workloads. While such heuristics capture certain domain insights, rules crafted for one system or version rarely transfer to others, forcing experts to rebuild strategies.

    \myline{ML-based Approaches} To overcome these limitations, researchers have proposed automated approaches that typically follow a two‑phase workflow. First, they collect representative workload characteristics (e.g., I/O patterns, query plan cost vectors) by automated profiling or user queries. And second, different optimization engines (like reinforcement‐learning (RL) agents using policy gradients) iteratively propose configurations that are tested on offline benchmarks before finalization. Prominent examples include CherryPick \cite{alipourfard_cherrypick_2017} and DBTune \cite{noauthor_pku-dairdbtune_2024}, which leverage Gaussian‑process–based Bayesian optimization to reduce tuning overhead by up to 75\% on cloud analytics workloads; QTune \cite{li_qtune_2019} and CDBTune \cite{zhang_hbox_2021}, apply deep RL-models to adapt configurations under dynamic query and cloud‑instance patterns; and OtterTune's \cite{van_aken_inquiry_2021} meta‑learning engine fingerprints workloads to transfer priors across deployments, cutting required trials nearly in half. While more general than heuristics-based methods, ML approaches are closely tied to training data and often lose effectiveness as workload, hardware, or software evolves. Further, they often focus on subsets of configurations, as with Endure \cite{Endure} and Dremel \cite{Dremel} for key-value store tuning. 

\vspace{-0.25em}
\section{Conclusion}

In this paper, we propose \sysname, a novel auto-tuning framework for heterogeneous storage systems that leverages specialized LLM agents to efficiently explore configurations while dynamically generating and managing tuning insights. This framework not only opens new opportunities for managing insights across storage systems but also introduces new measurement metrics for evaluating LLM-based configuration tuning systems.

\bibliographystyle{plainurl}
\bibliography{sample,zhichao-references,mai-references,gpt-references,llm-references}

\end{document}